\documentclass[12pt]{article}
\usepackage{psfig}
\usepackage{verbatim}
\usepackage{color}
\usepackage{rotate}
\usepackage{graphics}
\usepackage{epsf}
\usepackage{epsfig}
\usepackage{graphicx}
\usepackage{latexsym}
\usepackage{amscd}
\usepackage{amssymb}
\newcommand{\beq}{\begin{equation}}
\newcommand{\eeq}{\end{equation}}
\newcommand{\beqn}{\begin{eqnarray}}
\newcommand{\eeqn}{\end{eqnarray}}
\newcommand\noi{\noindent} 
\newcommand\la{\langle}
\newcommand\ra{\rangle}
\newcommand\eps\varepsilon
\newcommand\euler{{\rm e}}
\newcommand\imag{{\rm i}}

\def\fm{\,\mbox{fm}}
\def\GeV{\,\mbox{GeV}}

\def\lsim{\mathrel{\rlap{\lower4pt\hbox{\hskip1pt$\sim$}}
    \raise1pt\hbox{$<$}}}         
\def\gsim{\mathrel{\rlap{\lower4pt\hbox{\hskip1pt$\sim$}}
    \raise1pt\hbox{$>$}}}         

\begin{document}

\hfill {LA-UR-02-7759}

\vspace*{2cm}
\begin{center}
{\Large
\bf 
Relating different approaches to nuclear broadening}
\end{center}
\vspace{.5cm}

\begin{center}
 {\large
J\"org Raufeisen\footnote{\tt email: jorgr@lanl.gov}}\\
\medskip

{\sl Los Alamos National Laboratory,
Los Alamos, New Mexico 87545, USA}
\end{center}

\vspace{.5cm}

\begin{abstract}
\noi
Transverse momentum broadening of fast partons propagating through 
a large nucleus is proportional
to the average color field strength in the nucleus.
In this work, the corresponding coefficient is determined in  
three different frameworks, namely in
the color dipole approach, in
the approach of Baier et al.\ and in the higher twist factorization formalism.
This result enables one to use a
parametrization of the dipole cross section to estimate
the values of the gluon transport coefficient and of the higher twist 
matrix element, which is relevant for nuclear broadening. A considerable
energy dependence of these quantities is found. 
In addition,
numerical calculations are compared to data for nuclear broadening of 
Drell-Yan dileptons, $J/\psi$ and $\Upsilon$ mesons. 
The scale dependence of the strong coupling constant leads to
measurable differences
between the higher twist approach and the other two formalisms.
\medskip

\noi
\end{abstract}

\clearpage

\section{Introduction}
 
A fast parton (quark or gluon) propagating through nuclear matter
accumulates transverse momentum by multiple interactions
with the soft color 
field of the nucleus. At not too high energies, 
this
phenomenon is experimentally accessible by measuring
nuclear broadening of Drell-Yan (DY) dileptons or of $J/\psi$ and
$\Upsilon$ mesons produced in proton-nucleus ($pA$) collisions.
Nuclear broadening is defined as
the increase of the mean transverse momentum squared 
of the produced particle
compared to proton-proton ($pp$) collisions, i.e.\
\beq
\delta\la p^2_T\ra=\la p^2_T\ra_{pA}-\la p^2_T\ra_{pp}.
\eeq
During the past decade, at least three different theoretical
approaches 
have been developed to describe this effect, namely 
the color dipole approach \cite{dhk,jkt}, 
the approach of Baier et al.\ \cite{BDMPS}
and the higher twist factorization 
formalism \cite{Guo,HT} (see also \cite{early}
for earlier work).
This enormous interest in a QCD based description of nuclear effects
is mainly motivated by the experimental program at
the Relativistic Heavy Ion Collider (RHIC). 
Data from heavy ion collisions at RHIC require a profound 
theoretical understanding
of nuclear effects in terms of QCD for a reliable interpretation 
(see e.g.\ \cite{wang}).

In each of the three approaches \cite{dhk,jkt,BDMPS,Guo,HT}, 
broadening is proportional to a nonperturbative 
parameter, which has to be determined from experimental data.
This often limits the predictive power of the theory. Moreover,
one would like to know, how the different theoretical formulations
of transverse momentum broadening relate to one another, and to
what extend they represent the same (or different) physics. 
The purpose of this paper is to present relations between the nonperturbative
parameters, thereby illuminating the connection between these
seemingly very different approaches. 
Since the nonperturbative
input to the dipole approach \cite{dhk,jkt} is known from 
processes other than nuclear broadening, one can then obtain 
independent estimates for the parameters of the other
two approaches and calculate 
$\delta\la p^2_T\ra$ in a parameter free way. In addition, we study the
energy dependence of the nonperturbative parameters, which has
to be known if one wants to extrapolate results from fixed target
energies to RHIC and the Large Hadron Collider (LHC).

We shall now briefly summarize the basic formulae for nuclear broadening.
In the color dipole approach \cite{dhk,jkt}, transverse momentum
broadening of an energetic parton propagating through a large nucleus 
is given by
\beq\label{eq:dipole}
\delta\la p^2_T\ra^{\cal R}_{\rm dipole}=2\rho_A L C_{\cal R}(0),
\eeq
where $\rho_A=0.16\fm^{-3}$ is the nuclear density, and
 $L=3R_A/4$ is the average
length of the nuclear medium traversed by the projectile parton
before the hard reaction occurs ($R_A$ is the nuclear radius). 
The index ${\cal R}$ refers to the color representation of the projectile
parton, ${\cal R}={\cal F}$ for a quark and ${\cal R}={\cal A}$ for a gluon.
The nonperturbative physics is parametrized in the quantities
\beq\label{eq:cs}
C_{\cal F}(0)=\left.\frac{d}{dr_T^2}\sigma_{q\bar q}^N(r_T)\right|_{r_T\to0},
\quad
C_{\cal A}(0)=9C_{\cal F}(0)/4.
\eeq
Here, $\sigma_{q\bar q}^N(r_T)$ is the cross section for scattering a 
color singlet quark-antiquark ($q\bar q$) pair 
with transverse separation $r_T$
off a nucleon $N$. This dipole cross section arises from a complicated 
interplay between attenuation and multiple rescattering of the
incident parton \cite{jkt}.

In the approach of Baier et al.\ \cite{BDMPS} 
(BDMPS approach hereafter),
broadening is related to the transport
coefficient $\hat q_{\cal R}$,
\beq\label{eq:BDMPS}
\delta\la p^2_T\ra^{\cal R}_{\rm BDMPS}=\hat q_{\cal R}L.
\eeq
In this approach, all nonperturbative physics is contained in 
$\hat q_{\cal R}$, which 
is a measure for the strength of the interaction between the
projectile quark and the target. 

The dipole and the
BDMPS approach quite obviously describe the same 
physics, see Ref.~\cite{equiv}.
Both, $C_{\cal R}(0)$ and $\hat q_{\cal R}$ 
can be related to the gluon density 
of a nucleon. By comparing the corresponding expressions in 
Ref.~\cite{fsdcs} (for $C_{\cal R}(0)$) and Ref.~\cite{BDMPS}
(for $\hat q_{\cal R}$), one obtains \cite{urs2},
\beq
\hat q_{\cal R}=2\rho_AC_{\cal R}(0).
\eeq
Thus, $\delta\la p_T^2\ra^{\cal R}_{\rm dipole}
=\delta\la p_T^2\ra^{\cal R}_{\rm BDMPS}$.

\begin{figure}[h]
  \scalebox{1.0}{\includegraphics{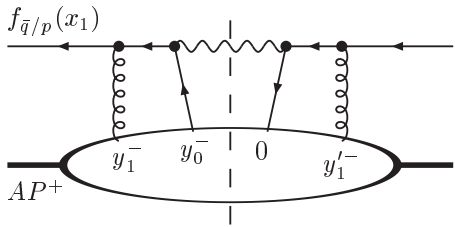}}\hfill
  \raise-0.5cm\hbox{\parbox[b]{3.44in}{
   \caption{\label{fig:ht}\em
Twist-4 contribution to nuclear broadening. 
The projectile antiquark carries momentum fraction $x_1$
of its parent hadron and undergoes one soft
rescattering before it annihilates with
a quark from the nucleus.
The Drell-Yan process is used to probe the transverse momentum
of the antiquark.
  	} 
  }
}
\end{figure}

The relation
to the higher twist factorization formalism \cite{Guo,HT}
is less clear.
In the dipole and in the BDMPS approach, 
the projectile parton acquires transverse
momentum in a random walk through the nuclear medium, thereby 
undergoing multiple soft rescatterings. In the higher twist approach, 
the (anti-)quark from the projectile proton 
exchanges only one additional soft
gluon
with the nucleus before the DY dilepton is produced, see Fig.~\ref{fig:ht}. 
Broadening then
depends on a particular twist-4 matrix element, which is
enhanced by a power of $A^{1/3}$ due to the size of the nucleus
($A$ is the
atomic mass of the nucleus) \cite{Guo},
\beq\label{eq:HT}
\delta\la p^2_T\ra^{\cal F}_{\rm HT}
=\frac{4\pi^2}{3}\alpha_s(M^2)\lambda_{{\rm LQS}}^2A^{1/3}.
\eeq
The quantity $\lambda^2_{{\rm LQS}}$ originates from a model of the soft-hard
twist-4 matrix element \cite{HT},
$T_{qG}^{SH}(x_2)\approx\lambda^2_{\rm LQS}A^{1/3}f_{q/A}(x_2),$
where $f_{q/A}(x_2)$ is the density of quarks with momentum fraction
$x_2$ in the nucleus. The strong coupling constant $\alpha_s$ enters at
the characteristic hard scale of the process that probes the transverse
momentum of the incident parton, i.e.\ the dilepton mass $M^2$. Thus,
in the higher twist approach, $\alpha_s$ is small, even though the exchanged 
gluon is soft. The smallness of $\alpha_s$ is crucial for the applicability 
of the QCD factorization theorem.

\section{Relating the dipole approach to the higher twist formalism}

In order to relate all three approaches, one clearly cannot simply
set equal the broadening given in
Eqs.~(\ref{eq:dipole}), (\ref{eq:BDMPS}) and (\ref{eq:HT}),
and then read off a relation between $C_{\cal R}(0)$, 
$\hat q_{\cal R}$ and
$\lambda^2_{{\rm LQS}}$.
Instead, one has to find a relation between 
these three quantities in an independent way,
and only after that, 
one can check whether all three approaches
predict the same (or different) $\delta\la p_T^2\ra^{\cal R}$. 
It is then possible to use a model for the dipole cross section to
estimate $\hat q_{\cal R}$
and $\lambda^2_{{\rm LQS}}$, since $\sigma^N_{q\bar q}$ is 
known much better than these
two quantities.

The plan is to relate 
$C_{\cal R}(0)$ and $\lambda^2_{{\rm LQS}}$ to the quantity
\beq\label{eq:meanf}
\la F^2\ra=\frac{1}{2\pi P^+}\int dy^-
\la N|F_a^{+\omega}(y^-)F_{a,\omega}^{+}(0)|N\ra,
\eeq
which measures the average color field strength experienced
by the projectile parton.  
In Eq.~(\ref{eq:meanf}), 
$P^+$ is the light-cone momentum of the nucleus $|A\ra$
per nucleon. 
The index $\omega$ runs over the two transverse directions,
and $F_a^{+\omega}$ is the gluon field strength operator.
Since we are dealing with non-perturbative
quantities, the result will of course be model dependent.

The relation between $C_{\cal R}(0)$ and $\la F^2\ra$ can be obtained 
quite straightforwardly.
Note
that the dipole cross section is related to 
the gluon density $xG_N(x)$ of a nucleon by \cite{fsdcs},
\beq
\sigma^N_{q\bar q}(x,r_T)=\frac{\pi^2}{3}\alpha_s r_T^2xG_N(x),
\eeq
and in light-cone gauge, the gluon density is given by \cite{fact},
\beq\label{eq:gdens}
xG_N(x)=
\int\frac{dy^-}{2\pi P^+} \euler^{-\imag xP^+y^-}
\la N|F_a^{+\omega}(y^-)F_{a,\omega}^{+}(0)|N\ra.
\eeq

What are the relevant scales for $xG_N(x)$ and $\alpha_s$?
Obviously, the Fourier modes of the nuclear color field that give the
dominant contribution to broadening are 
of order $\delta\la p_T^2\ra^{\cal R}$.
Since this is not much larger than 
$\Lambda_{QCD}^2$ in present experiments, the gluon density in this case
is not a parton distribution like in deeply inelastic scattering (DIS)
and should not be evolved with the QCD evolution equations. 
Moreover,
the soft gluon carries momentum fraction 
$x\sim\delta\la p_T^2\ra^{\cal R}/(x_1S)$ of its parent nucleon, 
which is essentially zero. Here, $x_1$ is
the momentum fraction of the projectile parton, and $S$ is the hadronic
center of mass (c.m.) energy.
We therefore set $x\to 0$ in
Eq.~(\ref{eq:gdens}) and write $C_{\cal F}(0)$ as
\beq\label{eq:c}
C_{\cal F}(0)=\frac{\pi^2}{3}\alpha_s(\delta\la p_T^2\ra^{\cal F})\la F^2\ra.
\eeq

It is important to note 
that in the dipole approach and in the BDMPS approach, the scale of
$\alpha_s$ is the same as in the gluon density. This is the main difference
between these two approaches and the higher twist formalism. 

The next step is to find an expression for $\lambda^2_{\rm LQS}$.
We shall follow
the model assumptions of Ref.~\cite{HT}, 
{\footnotesize
\beqn\label{eq:matrix}
\nonumber
\lefteqn{T_{qG}^{SH}(x_2)=}\\
&=&\int\frac{dy_0^-}{2\pi}\frac{dy_1^-dy_1^{\prime-}}{2\pi}
\euler^{\imag x_2 P^+y_0^-}\Theta(y_0^--y_1^-)\Theta(-y_1^{\prime-})
\frac{1}{2}\la A|\bar q(0)\gamma^+ q(y_0^-)
F_a^{+\omega}(y_1^-)F_{a,\omega}^{+}(y_1^{\prime-})|A\ra\\
\label{eq:zwei}&\approx&
\int\!\!\frac{dy_0^-}{2\pi}\euler^{\imag x_2 P^+y_0^-}
\frac{1}{2}\la A|\bar q(0)\gamma^+ q(y_0^-)|A\ra
\int\!\!\frac{dy_1^-dy_1^{\prime-}}{2\pi 2P^+V}
\Theta(y_0^--y_1^-)\Theta(-y_1^{\prime-})
\la A|F_a^{+\omega}(y_1^-)F_{a,\omega}^{+}(y_1^{\prime-})|A\ra\\
&\approx&
\lambda^2_{\rm LQS}A^{1/3}f_{q/A}(x_2).
\eeqn}
Here, $x_2$ is the momentum fraction of the quark from the nucleus in
Fig.~\ref{fig:ht}. The meaning of the positions $y_i^-$ on the light-cone are
illustrated in Fig.~\ref{fig:ht} as well.
The step functions
$\Theta$ ensure that the soft gluon is exchanged before 
the annihilation. In Eq.~(\ref{eq:zwei}),
the matrix element is factorized
by introducing an approximate unit operator, 
${\mathbf 1}\approx|A\ra\la A|/(2P^+V)$, where
$V$ is the volume of the nucleus \cite{r2}.
In this step, all
correlations between the quark and the gluon in  Fig.~\ref{fig:ht}
are neglected. As pointed out in Ref.~\cite{HT}, one has $|y_0^-|\ll R_A$
because of the rapidly oscillating phase factor in Eq.~(\ref{eq:zwei}).
In addition, $|y_1^--y_1^{\prime-}|\ll R_A$ because of 
confinement \cite{HT}. This allows one to approximate 
$\Theta(y_0^--y_1^-)\approx\Theta(-y_1^-)\approx\Theta(-y_1^{\prime-})$
in Eq.~(\ref{eq:zwei}).
With these approximations, the $y^-_0$-integral factorizes to give
the nuclear quark density $f_{q/A}(x_2)\approx A f_{q/N}(x_2)$.
The integral over the remaining step function yields a factor $L$,
and in the last integration one recovers the right-hand side of
Eq.~(\ref{eq:meanf}), though with $|N\ra$ replaced by $|A\ra$. 
Assuming that there are no non-trivial
nuclear effects on the gluon field,
the result reads,
\beq\label{eq:lam}
\lambda^2_{\rm LQS}A^{1/3}=\frac{1}{2}\rho_AL\la F^2\ra.
\eeq
Note that we do not introduce
a new model for $T^{SH}_{qG}(x_2)$. Eq.~(\ref{eq:lam}) follows
from the model assumptions of Ref.~\cite{HT}.

Thus, in all three approaches, broadening is related to the quantity
$\la F^2\ra$, and one finds from
Eqs.~(\ref{eq:dipole}), (\ref{eq:BDMPS}) and (\ref{eq:HT}),
\beq\label{eq:result}
\delta\la p_T^2\ra^{\cal F}_{\rm dipole}=
\delta\la p_T^2\ra^{\cal F}_{\rm BDMPS}=
\frac{2\pi^2}{3}\alpha_s(\delta\la p_T^2\ra^{\cal F})\rho_AL\la F^2\ra
\quad,\quad
\delta\la p_T^2\ra^{\cal F}_{\rm HT}=
\frac{2\pi^2}{3}\alpha_s(M^2)\rho_AL\la F^2\ra.
\eeq
The new result here is the coefficient $2\pi^2\alpha_s\rho_AL/3$,
the proportionality between broadening and the average color field 
strength in the target was already known before \cite{HT,urs1,urs3}.
It is remarkable, that
the only difference between $\delta\la p_T^2\ra^{\cal F}_{\rm dipole}$ and 
$\delta\la p_T^2\ra^{\cal F}_{\rm HT}$ is the scale of the strong coupling 
constant. We stress that 
this difference cannot be dismissed as a higher order 
correction. Instead, it is the result of different physical pictures
of nuclear broadening. 

At first sight, the result Eq.~(\ref{eq:result}) may seem puzzling. How can 
the double scattering approximation yield essentially the same expression
for broadening
as a resummation of all rescatterings? In fact, it was demonstrated in
Refs.~\cite{jkt,urs3}, that double scattering 
does not lead to an $A^{1/3}$-dependence of broadening.

This contradiction can be resolved
in the following way: The probability to
have $n$ interactions of the projectile parton with the medium
before the Drell-Yan process takes place is (neglecting correlations)
Poisson distributed,
$P_n={\left(\sigma T_A\right)^n}\euler^{-\sigma T_A}/n!,$
where $\sigma$ is the cross section for a single soft scattering,
and $T_A$ is the nuclear thickness. Apparently, the 
$A$-dependence of the single scattering probability is quite different
from $A^{1/3}$. In the dipole and the BDMPS approach, the accumulated
transverse momentum is proportional to the mean number of scatterings, i.e.\
$\sigma T_A$, and hence proportional to $A^{1/3}$. 
Therefore, it was concluded in \cite{jkt} that it is essential 
to sum all rescatterings in order to get an  $A^{1/3}$ law.
The higher twist approach, however, does not only use the double scattering
approximation, it is also an expansion in $\sigma T_A$. To leading order in
this parameter, $P_1$ is identical to the mean number of rescatterings.
This property of the Poisson distribution is the reason why the two
expressions for $\delta\la p_T^2\ra^{\cal F}$ in Eq.~(\ref{eq:result})
can be so similar.
In fact, it has been shown
recently \cite{rainer} that Eq.~(\ref{eq:HT}) remains valid, 
if the projectile
quark exchanges an arbitrary number of gluons with the target nucleus.
It should be stressed at this point, that 
$\delta\la p_T^2\ra^{\cal R}$ only depends on the average color field strength
in the target and is not sensitive to details of the color field.
Regarding details of the $p_T$ dependence of nuclear effects, one certainly 
has to expect very different expressions from the dipole approach and
the higher twist formalism.

\section{Phenomenological applications}

One can now choose a particular model of the dipole cross section to get 
an estimate for $\hat q_{\cal R}$ and $\lambda_{\rm LQS}^2$. In
this paper, the 
parametrization of Kopeliovich, Sch\"afer and Tarasov (KST) \cite{KST}
will be used,
because it is motivated from the phenomenology of soft hadronic interactions.
With the KST-parametrization, 
$C_{\cal R}(0)=C_{\cal R}(0,s)$ depends on the energy
$E_p$ of the projectile parton, $s=2m_NE_p$, where $m_N$ is the nucleon mass.
In all calculation, we also take into account Gribov's inelastic corrections
(i.e.\ gluon shadowing),
as explained in \cite{jkt}. 
At fixed target energies, 
this leads only to a $\sim 10\%$ reduction of $C_{\cal F}(0,s)$
for a heavy nucleus, but at larger values of $\sqrt{s}$, which are relevant 
for LHC, $C_{\cal F}(0,s)$ is reduced by 
approximately $1/3$.

\begin{figure}[t]
  \scalebox{0.4}{\includegraphics{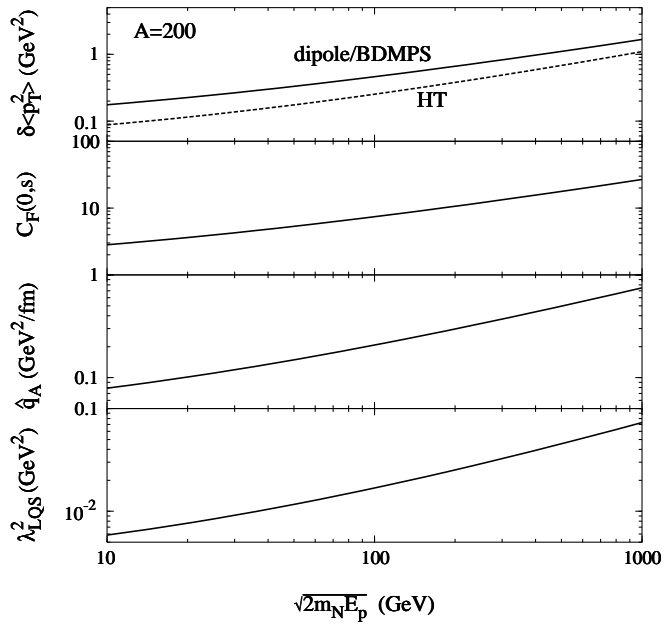}}\hfill
  \raise1.5cm\hbox{\parbox[b]{2.44in}{
   \caption{\label{fig:par}\em
Upper panel: Broadening for a quark
propagating through a large nucleus
as function of $\sqrt{s}=\sqrt{2m_NE_p}$.
Because of the hard scale in $\alpha_s$, broadening
is smaller in the higher twist ($HT$)
approach (dashed curve) than in the dipole and the BDMPS approach
(solid curve). The dashed curve is calculated from 
Eqs.~(\ref{eq:HT},\ref{eq:lqs})
with scale $M=5\GeV$ in
$\alpha_s$.
The other three panels show the energy dependence of the 
nonperturbative parameters of each approach.
  	} 
  }
}
\end{figure}

Fig.~\ref{fig:par} shows the energy dependence 
of nuclear broadening for quarks and 
of the three parameters
$C_{\cal F}(0,s)$, $\hat q_{\cal A}=9\rho_A C_{\cal F}(0,s)/2$ and
\beq\label{eq:lqs}
\lambda_{\rm LQS}^2A^{1/3}
=\frac{3}{4\pi^2\alpha_s(\delta\la p_T^2\ra^{\cal F})}
\la T_A \ra
C_{\cal F}(0,s),
\eeq
where $\la T_A \ra=\int d^2b T^2_A(b)/A$ is the nuclear thickness function
averaged over impact parameter $b$, and the strong coupling constant is
evaluated at a scale 
$\delta\la p_T^2\ra^{\cal F}=\la T_A \ra C_{\cal F}(0,s)$. Since this
scale is in most cases too small for perturbative QCD, 
we use a running coupling constant that freezes at low scales
\beq\label{eq:alphas}
\alpha_s(Q^2)=\frac{4\pi}{9\ln
\left(\frac{Q^2+0.54 {\rm GeV}^2}{0.04{\rm GeV}^2}\right)},
\eeq
in the spirit of \cite{dok}. In all three approaches, 
broadening has a quite significant energy dependence,
which is due to gluon radiation included in the KST-parametrization.

We obtain a value of $\lambda_{\rm LQS}^2\approx 0.008$
at $\sqrt{s}=22\GeV$ (which is the quark energy relevant for 
Fermilab fixed target kinematics) 
that is very close the one of 
Ref.~\cite{Guo} ($\lambda_{\rm LQS}^2=0.01\GeV^2$), see
Fig.~\ref{fig:par}. The latter value was determined in Ref.~\cite{Guo} 
from E772 data on broadening for DY.
For the gluon transport coefficient, one obtains  
$\hat q_{\cal A}(\sqrt{s}=22\GeV)\approx0.11\GeV^2/\fm$, 
more than $2$ times as large as the one estimated in Ref.~\cite{jhep} 
($\hat q_{\cal A}\approx0.045\GeV^2/\fm$).

In the dipole approach, broadening only depends on the
energy of the parton and not on the mass of the dilepton.
In the higher twist formalism, however, $\delta\la p_T^2\ra^{\cal F}$ 
depends on the dilepton mass through $\alpha_s$.
As a consequence, 
for $W^\pm$ and $Z^0$ production in $pA$ scattering
with $\sqrt{S}=8.8$ TeV at the LHC 
($x_1\approx x_2\approx 0.01$), 
one has
$\delta\la p_T^2\ra^{\cal F}_{\rm dipole}\sim 1.5\GeV^2$ for a heavy 
nucleus with $A\sim200$ and 
$\delta\la p_T^2\ra^{\cal F}_{\rm HT}\sim 0.5\GeV^2$.  
Of course,
this estimate assumes that one can still apply these
formalisms at $x_2=0.01$.
As explained in more detail in
Ref.~\cite{krtj}, at very low $x_2$, the DY cross section is affected by
quantum mechanical interferences, and the transverse momentum
broadening of the produced boson does not reflect the broadening
of the projectile quark any more. Nuclear broadening in DY 
at very low $x_2$ has been calculated in Ref.~\cite{krtj} and
is expected to be much larger than at medium-low $x_2\gsim0.01$.   

\begin{figure}[ht]
  \centerline{\scalebox{0.45}{\includegraphics{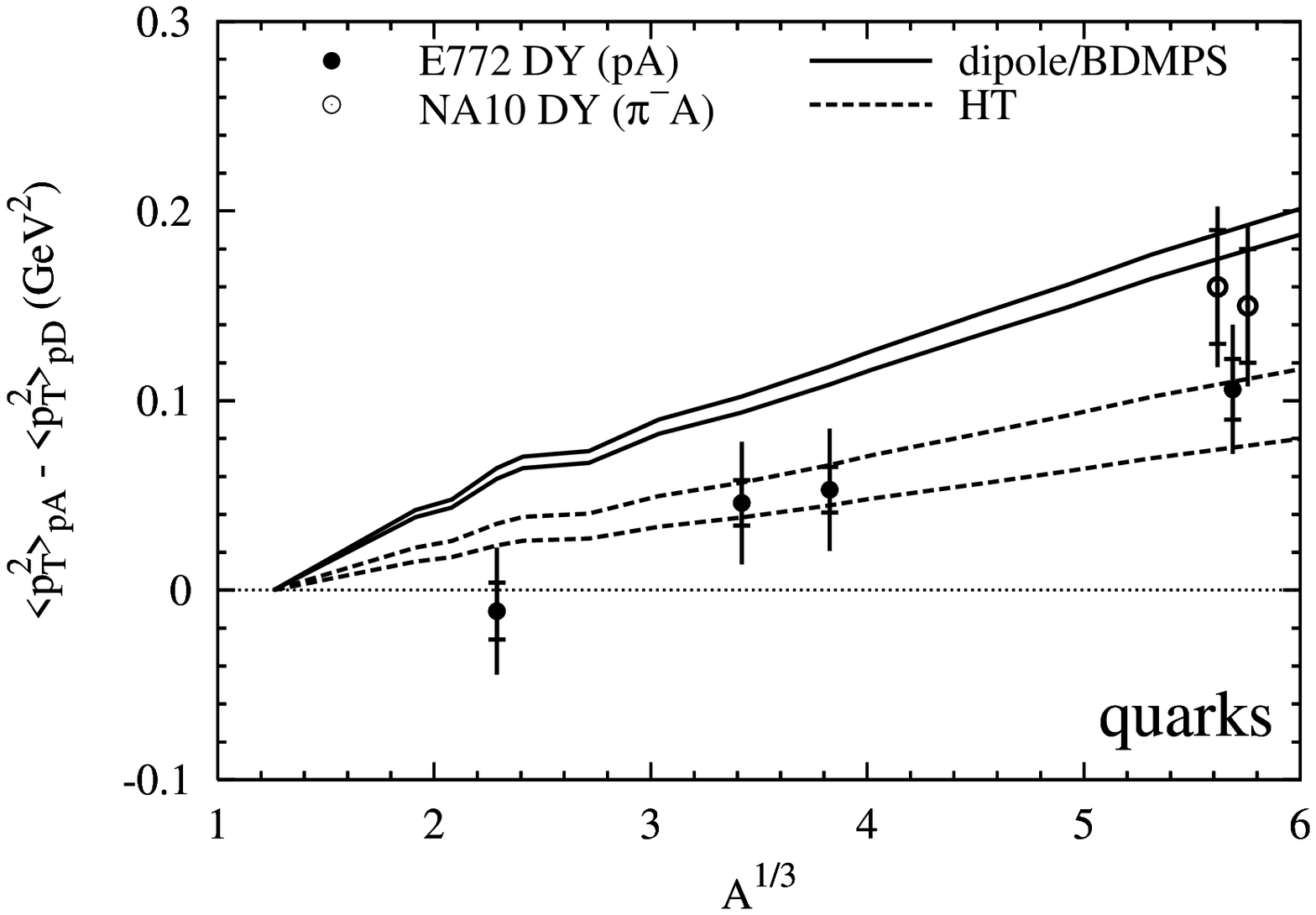}}\hfill
	      \scalebox{0.45}{\includegraphics{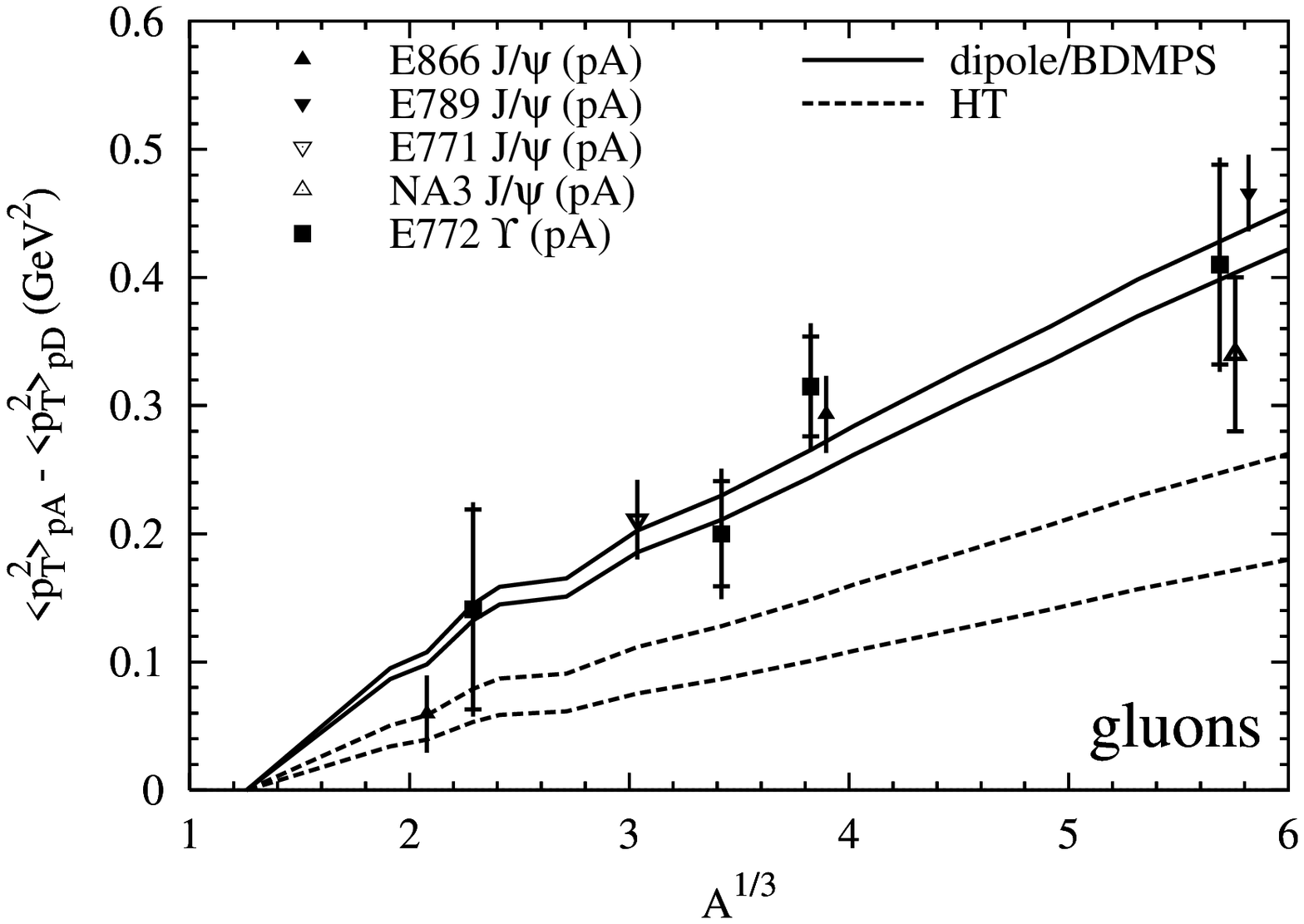}}}
    \center{\parbox[b]{16cm}{\caption{
      \label{fig:broad}\em
Calculations vs.\ experimental data 
\cite{E772,NA10,E866} for broadening with respect to (w.r.t.)
deuterium ($D$).
Only the NA3 point \cite{NA3} represents broadening w.r.t.\ a proton.
The space in between the curves is an estimate of the theoretical
uncertainty. Inner error bars show statistical errors, outer
errorbars statistical and systematic errors added in quadrature.
Some points have been slightly displaced for better visibility.
The beam energies are $800\GeV$ (E$xxx$), $140\GeV$ and $286\GeV$ (NA10)
and $200\GeV$ (NA3). }  
    }  }
\end{figure}

At fixed target energies, however, these interference effects are 
negligible, and experimental data for broadening in DY can be 
compared to a calculation of broadening for quarks.
The solid curves in Fig.~\ref{fig:broad}
are obtained from
\beq\label{eq:corr}
\delta\la p_T^2\ra^{\cal R}_{pA}-\delta\la p_T^2\ra^{\cal R}_{pD}=
\left(\la T_A \ra -\la T_D \ra\right) C_{\cal R}(0,s),
\eeq
where the mean nuclear thickness is calculated with realistic 
parametrizations of nuclear densities from Ref.~\cite{dejager}.
The dashed curves in Fig.~\ref{fig:broad} are obtained by rescaling 
$\delta\la p_T^2\ra^{\cal R}_{pA}$ by the ratio of strong coupling constants,
$\alpha_s(M^2)/\alpha_s(\delta\la p_T^2\ra^{\cal R}_{pA})$.
The higher twist formalism is strictly speaking not applicable
to light nuclei, since all contributions that are not enhanced by a power
of $A^{1/3}$ are neglected in this approach. Nevertheless, we believe 
that a calculation with realistic nuclear densities is a reasonable 
extrapolation to lighter nuclei.

The relevant quark energies for the $800\GeV$ proton beam at Fermilab
are $20\GeV\le\sqrt{s}\le25\GeV$.
The lower solid and dashed curves
in Fig.~\ref{fig:broad} are calculated for $\sqrt{s}=20\GeV$ and
the upper ones for $\sqrt{s}=25\GeV$.  For the
higher twist calculation, we vary the scale of $\alpha_s$ 
in between the $J/\psi$ and the $\Upsilon$ mass.
This may serve as an estimate of the
theoretical uncertainty. As already noted in Ref.~\cite{jkt}, 
the dipole approach overestimates the DY data from E772 \cite{E772} by  
several standard deviations. This large discrepancy cannot
be explained by uncertainties in the parameterization of
the dipole cross section \cite{jkt}. 

However, we point out that the 
E772 values for broadening \cite{E772} were extracted only from DY data
with transverse momentum $p_T\lsim 3\GeV$ \cite{Pat}, where the
$p_T$-differential DY cross section is still nuclear enhanced, and may
therefore underestimate the true value of $\delta\la p_T^2\ra^{\cal F}$.
Moreover, the ${\cal O}(\alpha_s)$ parton model does not describe 
some of 
the $p_T$-integrated DY cross sections measured by E772, either
\cite{rpn}.
A future analysis \cite{jkm} based on new E866 data \cite{webb}, 
will include DY data with transverse momentum up to $p_T\lsim 5\GeV$,
and may yield values of $\delta\la p_T^2\ra^{\cal F}$ 
that are twice as large \cite{boris}.
One can therefore regard the curves
in Fig.~\ref{fig:broad} as predictions.

It is interesting to note that, while E772 only used dileptons
with  $p_T\lsim 3\GeV$,  the transverse momentum imbalance in
photoproduction of dijets was measured by
E683 \cite{E683} only for
jets with $p_T>3\GeV$. It has been argued in \cite{b2}, 
that the unusually large effect observed by E683 is (in part)
caused by this
restriction on $p_T$. In fact, a value of 
$\lambda^2_{\rm LQS}\approx 0.1 \GeV^2$ is needed to accommodate the E683
result \cite{HT}. The analysis presented in this paper clearly favors a much 
lower value, which is more consistent with the DY data.

The calculations in the dipole approach for  
broadening of gluons agree quite well with $J/\psi$ and $\Upsilon$ data,
which are underestimated by the higher twist formalism,
see Fig.~\ref{fig:broad} (right).
Of course, broadening for gluons is equal to broadening in 
$J/\psi$ and $\Upsilon$ production, only if final state effects 
are negligible. 
This assumption is
justified by the observation that broadening is very similar
(within errorbars) for $J/\psi$ and $\Upsilon$ mesons.

\section{Summary}

In this paper, we quantitatively related the color dipole approach 
\cite{dhk,jkt} to the
higher twist factorization formalism \cite{Guo,HT}, and studied
transverse momentum broadening of fast partons propagating through
cold nuclear matter.
In both approaches, broadening is proportional
to the average color field strength experienced by the projectile parton
\cite{HT,urs1,urs3}.
We find that the corresponding coefficients differ only by the scale of the 
strong coupling constant. While broadening is an entirely soft process
in the dipole approach, the extension of the QCD factorization theorem to
twist-4 is justified by the smallness of $\alpha_s$. In the higher twist
formalism, $\alpha_s$ enters at the
typical hard scale of the process that probes the transverse momentum
of the projectile parton.
The equivalence between the dipole and the BDMPS approach \cite{BDMPS}
was already known 
before \cite{equiv}. 

Since the dipole cross section is much better constrained by data
than $\lambda^2_{\rm LQS}$ and $\hat q_{\cal R}$, one is now able to obtain
new estimates for the latter two quantities. 
So far, $\lambda^2_{\rm LQS}$ could be determined only from the same
data the higher twist approach is supposed to describe
\cite{Guo},
and estimates for $\hat q_{\cal R}$ were based mostly on physical 
intuition \cite{jhep}. 
With the KST-parameterization of the dipole cross section \cite{KST}, 
which we use, 
 broadening is 
a function of the energy of the projectile parton, as one would expect 
from a soft process. In the higher twist approach, there is an additional
scale dependence through $\alpha_s$.
To our best knowledge, this is the first time that
quantitative results for the energy dependence of 
$\lambda^2_{\rm LQS}$ and $\hat q_{\cal R}$ are presented. It will
be necessary to take this energy dependence into account, when 
applying the higher twist formalism and the BDMPS approach at RHIC
or even at LHC energies. 

At fixed target energies, 
numerical calculations in the dipole approach exceed results obtained in
the higher twist formalism by a factor of $\sim2$. Most importantly,
the uncertainty bands of both approaches do not overlap, if
one varies the remaining free parameters within reasonable limits.
Available data, however, do not yet allow to rule out one of the theories.
Though the dipole approach describes $J/\psi$ and $\Upsilon$ data well, 
this agreement has to be interpreted with 
great care, since final state effects are not taken into account by the
theory. We argue, however, that the similarity between 
broadening for $J/\psi$ and $\Upsilon$ mesons indicates that final state
effects are rather small. The higher twist approach underestimates 
broadening for $J/\psi$ and $\Upsilon$ mesons.
Broadening for DY, on the other hand, is overestimated  
in the dipole approach, while the higher twist formalism
reproduces these data well. However, the small values
of $\delta\la p_T^2\ra^{\cal F}$ measured by E772 may be the result of a
too low $p_T$ cut imposed on the data. 
A reevaluation of the E772 data in question, as
well as new results from E866 measurements, are expected soon \cite{jkm}.
This new analysis
will probably yield significantly larger broadening for DY dileptons
\cite{Pat,boris}. We stress that no parameter in our calculations has been
adjusted to fit the data. Thus, the curves presented here can
be regarded as predictions.

\medskip
\noindent {\bf Acknowledgments}: 
I am indebted to Rainer Fries, Mikkel Johnson, Boris Kopeliovich,
Pat McGaughey and
Joel Moss for valuable discussion
and to Mike Leitch for providing the $J/\psi$-broadening data.
This work was supported by the
U.S.~Department of Energy at Los Alamos National Laboratory under Contract
No.~W-7405-ENG-38.

\end{document}